\documentclass[10pt,preprint,floatfix]{revtex4}
\usepackage[latin1]{inputenc}
\usepackage{graphicx}% Include figure files
\usepackage{dcolumn}% Align table columns on decimal point
\usepackage{bm}% bold math t
\usepackage{amssymb,amsmath}
\usepackage{hyperref}
\usepackage{hyphenat}
\usepackage{url}

\begin{document}

%==========================================================
\newcommand{\be}{\begin{equation}}
\newcommand{\ee}{\end{equation}}
\newcommand{\ben}{\begin{eqnarray}}
\newcommand{\een}{\end{eqnarray}}
\newcommand{\nn}{\nonumber \\}
\newcommand{\ii}{\'{\i}}
\newcommand{\pp}{\prime}
\newcommand{\tr}{{\mathrm{Tr}}\,}
\newcommand{\nd}{\noindent}
\newcommand{\nl}{\newline}

%==========================================================
% Front Matter (Article Type, Title, Authors, etc.)
%---------------------------------------------------------

% Full title of the paper
\title{Quantum echoes in classical and semiclassical statistical treatments}

% Authors
\author{F. Pennini$^{1,2}$, A. Plastino$^{2,3,4}$}

% Affiliations / Addresses
\affiliation{$^{1}$ Departamento de F\'{\i}sica, Universidad Cat\'olica del Norte, Av. Angamos 0610, Antofagasta, Chile\\
$^{2}$ Instituto de F\'{\i}sica La Plata--CCT-CONICET, Fac. de Ciencias Exactas,
 Universidad Nacional de La Plata, C.C.~727, 1900 La Plata, Argentina.\\ $^3$Departamento de Física and IFISC, Universitat de les Illes Balears, 07122 Palma de Mallorca, Spain.\\$^{4}$Instituto Carlos I de Física Teórica y Computacional and
Departamento de Física Atómica, Molecular y Nuclear, Universidad de Granada, Granada, Spain.}

\date{\today}

% Abstract
\begin{abstract}

Some quantal systems require only a small part of the full quantum
theory for their analysis in classical terms. In such
understanding we review some recent literature on semiclassical
treatments.
  An analysis of it allows one to  see that some important quantum features of the
harmonic oscillator can indeed  be already encountered at the classical or
semiclassical  statistical levels.

\end{abstract}

%\pacs{03.65.Sq, 03.67.-a, 05.30.-d, 42.50.Lc}

%\keywords{ information theory; phase space; semiclassical
%information; delocalization; purity; Fisher
%measures}

\maketitle

%%%%%%%%%%%%%%%%%%%%%%%%%%%%%%%%%%%%%%%%%%%\enddocument
%==========================================================

%==========================================================
% Main Text of the Paper
%----------------------------------------------------------

 %%%%%%%%%%%%%%%%%%%%%%%%%%%%%%%%%%%%%%%%%%%%%%%%%%%%%%%%%%%%%%%
\section{Introduction}
 %%%%%%%%%%%%%%%%%%%%%%%%%%%%%%%%%%%%%%%%%%%%%%%%%%%%%%%%%%%%%%%

It has been pointed out long ago that some quantal systems
require only a small part of the full quantum theory for their
analysis in classical terms
\cite{boyer}.
    With this notion in mind we review some recent work  \cite{PPFMandel,PPFMandel2,ourreview,Brazilian,unsere,Pennini1,PPFano}.
 After inspection, reflection,  and re-elaboration it will become apparent that some
typical quantal peculiarities can be explained, to a rather surprising extent, by recourse to just classical or
semiclassical considerations. We have in mind here such ``purely-quantum'' concepts as those of
 decoherence factor and Mandel parameter. We will
encounter quantum echoes regarding such notions, outside the Schröedinger or Heisenberg treatments. Our main research tools
 will be  escort
distributions, intertwined with
information-quantifiers, of which
 the semiclassical Wehrl's entropy and Fisher's
information measure are to be employed.

 The harmonic oscillator (HO) constitute the focus of our attention. This is, of course,  much more than a mere example,
  since in addition to the extensively used Glauber states in molecular physics and chemistry
   \cite{Klauder,molecular},  nowadays the HO
is of particular interest for the dynamics of bosonic or fermionic
atoms contained in magnetic traps \cite{davis,bradley}, as well as
for any system that exhibits an equidistant level spacing in the
vicinity of the ground state, like nuclei or Luttinger liquids. We briefly review below the notions undelying this communication.

%%%%%%%%%%%%%%%%%%%%%%%%%%%%%%%%%%%%%%%%%%%%
\subsection{Escort distributions}
%%%%%%%%%%%%%%%%%%%%%%%%%%%%%%%%%%%%%%%%%%

%%%%%%%%%%%%%%%%%%%%%%%%%%%%%%%%%%%%%%%%%%%%
 Given a probability distribution (PD) $f(x)$, there exists an
infinite family of associated PDs $f_{q}(x)$  given~by
 \be \label{launo}  f_q(x)= \frac{f^{q}(x)}{\int \mathrm{d}x\,
f^{q}(x)},\ee with $q$ a real parameter,  that  have proved to
be quite useful in the investigation of nonlinear dynamical
systems, as they often are better able to discern some of the
system's features than the original distribution~\cite{beck}. Here
we will take advantage of the $q-$degree of freedom to look for
effects not visible at $q=1$ that hopefully emerge at other
$q-$values. Additionally, it will be seen that physical
considerations constrain the $q-$choice.

\subsection{Decoherence}

Decoherence is that interesting  process whereby the quantum
 mechanical state of any macroscopic system becomes rapidly
correlated with that of its environment in such a manner that no
measurement on the system alone (without a simultaneous
measurement of the complete state of the environment) can exhibit
any interference between two quantum states of the system.
Decoherence is a rather exciting  phenomenon and a subject of
 widespread attention~\cite{tano}.
%%%%%%%%%%%%%%%%%%%%%%%%%%%%%%%%%%%%%%%%%%%%%%%%%%%%%%%%%%%%%%%%%%%%%%%%%%%%%%%%
 However, it is difficult
to provide a quantitative definition of it. All pertinent attempts
 always depend on the relevant experimental configuration and on the
authors' taste~\cite{tano2}.
%%%%%%%%%%%%%%%%%%%%%%%%%%%%%%%%%%%%%%%
An important related quantity is the square of the density matrix,
in whose terms one can define a decoherence parameter $\mathcal{D}$
\cite{watanabe}, ranging between 0 (pure states) and one. It is
defined as \be \label{laD} \mathcal{D}=1-\frac{\tr(\hat
\rho^2)}{(\tr \hat \rho)^2}. \ee This is a clearly non-negative
quantity.
 The quantity  $\tr(\hat
\rho^2)$ is often called the purity of $\hat \rho$, equal to unity for pure states.
\subsection{Mandel parameter and Fano factor}
\nd A convenient noise-indicator of a non-classical field is the
so-called Mandel parameter which is defined by~\cite{Mandel}

\be \label{afano} \mathcal{Q}=\frac{(\Delta \hat N)^2}{\langle \hat N\rangle}-1
\equiv \mathcal{F}-1, \ee which is closely related to the normalized
variance (also called the quantum Fano factor $\mathcal{F}$ \cite{Bajer})
$\mathcal{F}=(\Delta\hat N)^2/\langle\hat N\rangle$ of the photon
distribution. For $\mathcal{F} < 1$ ($\mathcal{Q} \le 0$), emitted light is referred
to as sub-Poissonian since it has photo-count noise smaller than
that of coherent (ideal laser) light with the same intensity
($\mathcal{F}=1;\,\,\mathcal{Q}=0$), whereas for $\mathcal{F} > 1$, ($\mathcal{Q} > 0$) the light is
called super-Poissonian, exhibiting photo-count noise higher than
the coherent-light noise.
 Of course, one wishes to minimize the
Fano factor.
%%%%%%%%%%%%%%%%%%%%%%%%%%%%%%%%%%%%%%%%%%%%%

%%%%%%%%%%%%%%%%%%%%%%%%%%
\section{Basic tools}
%\subsection
\nd We introduce next the basic tools needed for our endeavor.

\subsection{Phase-space, coherent states, and Husimi distributions}

\nd  In phase-space, exact quantum solutions are given by Wigner
distributions \cite{PRD2753_93,qoptics,Tannor}.  The paradigmatic
semiclassical concept to be  appealed to is that of Husimi
probability distribution, $\mu(x,p)$, built upon using coherent
states \cite{Glauber,Husimi,ourreview}. The pertinent
definition reads

\be \mu(x,p)=\langle z|\hat \rho|z\rangle,\label{i1DOS}\ee
     a ``semi-classical'' phase-space distribution function
 associated to the density matrix $\hat\rho$ of the
 system~\cite{Glauber,Klauder}.  Coherent states are eigenstates of the
 annihilation operator $\hat a$, i.e., satisfy $\hat a \vert z \rangle=z\vert z \rangle$.
 The distribution  $\mu(x,p)$ is
   normalized in the fashion \be \int \frac{\mathrm{d}x\, \mathrm{d}p}{2 \pi \hbar}\,\mu(x,p)=1. \ee  Indeed,   $\mu(x,p)$ is a
 Wigner-distribution $D_W$ smeared over an $\hbar$ sized region of phase space
 \cite{PRD2753_93}.
   The smearing renders $\mu(x,p)$ a positive function, even if $D_W$
  does not have
    such a character. The semi-classical Husimi probability
distribution
 refers to a special type of probability:
  that for simultaneous but approximate location of position and
 momentum in phase-space~\cite{PRD2753_93}.

 \nd  The usual treatment of equilibrium in statistical mechanics makes
 use of
 the Gibbs's canonical distribution,  whose
    associated, ``thermal''
     density matrix is given by  \be \hat\rho=Z^{-1}e^{-\beta \hat H}, \label{thermal}  \ee with $Z=\tr(e^{-\beta \hat H})$  the partition function,
 $\beta=1/k_BT$
     the inverse  temperature $T$,
    and $k_B$ the Boltzmann constant.

\subsection{Information quantifiers in phase-space}

\nd  The operative  semiclassical entropic measure is here Wehrls's entropy $W$, a  useful measure of localization in
phase-space~\cite{Wehrl}. Its definition reads

\be W=-\int \frac{\mathrm{d}x\, \mathrm{d}p}{2 \pi \hbar}\,
 \mu(x,p)\, \ln \mu(x,p).\label{i1}\ee The uncertainty principle manifests itself  through the inequality
\be \label{liebbound}
 1 \leq W,\ee which was first conjectured by
Wehrl~\cite{Wehrl} and later proved by Lieb~\cite{Lieb}.
In order to conveniently write down an expression for
    $W$ consider an arbitrary Hamiltonian $\hat H$ of eigen-energies $E_n$
     and eigenstates   $|n\rangle$ ($n$ stands for a collection of all the pertinent
     quantum numbers required to label the states).
 One can
  always write~\cite{PRD2753_93}
 \be \mu(x,p)=\frac{1}{Z} \sum_{n}
 e^{-\beta
   E_n}|\langle z|n\rangle|^2.\label{husimi0}
 \ee A useful route to $W$ starts then with Eq.~(\ref{husimi0}) and
 continues with Eq.~(\ref{i1}). In the special case of the harmonic oscillator the coherent states are of the form~\cite{Glauber}

 \be
 |z\rangle=e^{-|z|^2/2}\,\sum^{\infty}_{n=0}\frac{z^{n}}{\sqrt{n!}}\,|
    n\rangle, \label{qcoherent}
 \ee
where ${|n\rangle}$ are a complete orthonormal set of eigenstates
and whose spectrum of energy is $E_n=(n+1/2)\hbar\omega$,
$n=0,1,\ldots$ In this situation we have the useful analytic
expressions obtained in Ref.~\cite{PRD2753_93}

\be \mu(z)=(1-e^{-\beta\hbar\omega})\,e^{-(1-e^{-\beta\hbar\omega
})|z|^2}    \label{mu1},
 \ee
\begin{equation}\label{wehrl}
    W_{HO}=1-\ln(1-e^{-\beta\hbar\omega}).
\end{equation}
When $T\rightarrow 0$, the entropy takes its minimum value $W_{HO}=1$,
expressing purely quantum fluctuations. On the other hand when
$T\rightarrow \infty$, the entropy tends to the value
$-\ln(\beta\hbar\omega)$ which expresses purely thermal
fluctuations.
\vskip 2mm
\nd Fisher's information measure $I$ is the local counterpart of the  global Wehrl quantifier.
 It is an indicator of how much information is contained in a PDF \cite{frieden2}.
  In phase-space, the local quantifier adopts the appearance~\cite{Brazilian}
\be I=\frac14\, \int \frac{\mathrm{d}^2 z}{\pi}\,\mu(z)\,\left\{\frac{\partial \ln \mu(z)}{\partial
|z|}\right\}^2, \label{Fishz}\ee so that inserting the $\mu-$expression into the above
expression we obtain for the HO  the analytic form

\be I_{HO}=1-e^{-\beta \hbar\omega},\label{Fisher} \ee so that $0\leq I_{HO}\leq 1$.

\subsubsection{\bf A first observation}

\nd Introducing (\ref{Fisher}) into the Wehrl expression we find

\be \label{primera} W_{HO}= 1 -\ln{(I_{HO})}, \ee which together with the Lieb
inequality seems to be telling us that too much information might
be incompatible with the uncertainty principle. Closer inspection
shows, however, that the above expression is valid for any values
of either $\beta$ or $\omega$. We will return to this point later
on, in connection with escort distributions.

%%%%%%%%%%%%%%%%%%%%%%%%%%%%%%%%%%%%%%%%%%%%
\subsection{Escort Husimi distributions}
%%%%%%%%%%%%%%%%%%%%%%%%%%%%%%%%%%%%%%%%%%

%
\nd  Things can indeed be improved in the above described scenario by
recourse to this concept of escort distribution,  introducing
it in conjunction with
 semiclassical Husimi distributions. Thereby one might try to gather ``improved" {\it semiclassical
  information}  from escort  Husimi distributions ($q-$HDs)
$\gamma_q(x,p)$:

 \be \label{escortHus} \gamma_q(x,p)=
\frac{\mu(x,p)^{q}}{\int \frac{\mathrm{d}^2 z}{\pi}\,
    \mu(x,p)^q},\ee
    where $\mathrm{d}^2z/\pi=\mathrm{d}x\mathrm{d}p/2\pi\hbar$
and whose HO-analytic form can be obtained from Ref. \cite{unsere}, i.e.,
\be \gamma_q(z)=q (1-e^{-\beta\hbar\omega}) e^{-q (1-e^{-\beta\hbar\omega})|z|^2}. \label{DH escolta}\ee

\nd  As for the associated escort-Fisher measure $I_{sc}^{(q)}$ one easily gets
\begin{equation}\label{IF escolta}
    I_{sc}^{(q)}=\frac14\, \int \frac{\mathrm{d}^2 z}{\pi}\,\gamma_q(z)\,\left\{\frac{\partial \ln \gamma_q(z)}{\partial
|z|}\right\}^2,
\end{equation}
that using  (\ref{DH escolta}) leads to
\begin{equation}\label{IF escolta2}
    I_{sc}^{(q)}=q(1-e^{-\beta\hbar\omega})=q \,I_{HO},
\end{equation}
entailing that  $0<I_{sc}^{(q)} \leq q$.

\subsection{Coherent states and Mandel parameter}

\nd   For a coherent state (a pure quantum state) the
Mandel parameter vanishes, i.e., $\mathcal{Q} = 0$ and $\mathcal{F}=1$. A field in a
coherent state is considered to be the closest possible
quantum-state  to a classical field, since it saturates the
Heisenberg uncertainty relation and has the same uncertainty in
each quadrature component. It should be clear
that both $\mathcal{Q}$ and $\mathcal{F}$ function as indicators on non-classicality.
Indeed, for a thermal state one has $\mathcal{Q}>0$ and $\mathcal{F}>1$, corresponding
to a photon distribution broader  than the Poissonian. For $\mathcal{Q} <
0$, ($\mathcal{F}<1$) the photon distribution becomes narrower than that of
a Poisson-PDF and the associated state is non-classical. The most
elementary examples of non-classical states are number states.
Since they are eigenstates of the photon number operator $\hat N$,
the fluctuations in $\hat N$ vanish and the Mandel parameter reads
$\mathcal{Q} = -1$ ($\mathcal{F}=0$) \cite{qoptics}.
 Taking into account that the number operator is connected with
the harmonic oscillator Hamiltonian $\hat H$ via $\hat N=\hat
H/\hbar\omega-1/2$, we can rewrite the HO-Mandel parameter in this
fashion\be \mathcal{Q}=\mathcal{F}-1=\frac{(\Delta \hat H)^2}{\hbar\omega\langle \hat
H\rangle-\hbar^2\omega^2/2}-1,\label{Qmandel}\ee where we have
used that $ \hat H =\hbar\omega|z|^2$~\cite{Pennini1}.
 Of course, classically the hamiltonian phase-space function~is

 \be \label{classH} \mathcal{H}(x,p) = \frac{p^2}{2m} + \frac{1}{2}m\omega^2x^2. \ee

\section{Decoherence parameter}

\nd We shall calculate

\be \label{1laD} \mathcal{D}=1-\frac{\tr(\hat
\rho^2)}{(\tr \hat \rho)^2}, \ee
 it in three different versions: quantum,
classical, and semiclassical. In the two last instances, one replaces  $\hat\rho$ by an
ordinary, normalized PDF $f$ and the trace operation by integration over phase space, i.e.,

  \be \label{2laD} \mathcal{D}=1-\frac{\int\,\frac{\mathrm{d}x\mathrm{d}p}{h}\,f^2
}{(\int\,\frac{\mathrm{d}x\mathrm{d}p}{h}\,f)^2}. \ee Classically,
$\mathcal{D}$ is not guaranteed to be of a nonnegative character.
Interesting physical results ensue if we nonetheless demand
nonnegativity, as we shall see below.

\subsection{Quantal HO-version}

\nd  We begin with the orthodox quantum recipe. All our
calculations are performed in phase-space.
  For technical details consult, for instance, Ref. \cite{ourreview}. The quantum HO- density
  operator is $\hat\rho=e^{-\beta \hat H}/Z$, $\hat H$ the
 HO-Hamiltonian, and $Z=e^{-\beta\hbar\omega/2}/(1-e^{-\beta\hbar\omega})$ the partition function of this system, so that one straightforwardly finds

\be \mathcal{D}_{quant}= \frac{2}{1+e^{\beta\hbar\omega}}. \ee

\nd  It is easy to
see that for $\beta \rightarrow \infty$ one has
$\mathcal{D}_{quant}=0$ while for  $\beta \rightarrow 0$ one has
$\mathcal{D}_{quant}=1,$ as expected. Interesting things may
happen if we try to replace $\hat\rho$ by a classical PDF and the trace operation by integration over phase-space.

\subsection{Classical HO-version}

\nd Classically (or semiclassically), the delocalization factor
can be gotten by using probability distributions instead of
density matrices \cite{ourreview}. For the HO one has \be
\mathcal{D}_{class}=1-\frac{1}{Z_{class}^2}\int\frac{\mathrm{d}x\mathrm{d}
p} {h}e^{-2\beta\hbar\omega |z|^2}, \ee where $Z_{class}=1/(\beta\hbar\omega)$ is the classical partition function for the HO.
The pertinent computation  yields
\be \mathcal{D}_{class}=1-\frac{\beta\hbar\omega}{2}.
\label{laZ}\ee Interestingly enough, $\mathcal{D}_{class}
\rightarrow 1$ as $T \rightarrow \infty$, as in the quantum
instance.

\subsubsection{\bf First quantum echo}

\nd When dealing with Gaussian distributions one finds
$\mathcal{D}_{class} \ge 0$ [Cf. Ref. (\ref{laD})] only in special
cases.
 For $f=A e^{-a |z|^2}$ one readily finds \be \mathcal{D}_{class}=1-\frac{a}{2}.\ee
 Thus, $\mathcal{D}_{class}\geq 0$ implies $a\leq 2$. In our case, $a=\beta\hbar\omega$ and the requirement turns out
  to be that the ``thermal''
energy $k_BT$, i.e., the average classical energy per degree of
freedom $\langle e\rangle$, is such that \be \label{Dclass} \langle e\rangle_{min} \ge
\frac{\hbar\omega}{2}. \ee
 This  entails a rather surprising result,    {\it a
minimum possible mean energy per degree of freedom}
 $\langle e\rangle_{min}$. For energies smaller of this value the quantity
(\ref{2laD}) becomes negative. Thus, we encounter a
quantum-flavored result  at the classical level. One might  be
tempted to suggest that the vacuum energy $\hbar\omega/2$ has a
statistical origin. Why? Because a minimum possible HO-energy arises just by demanding that the pertinent distribution $f$ verify

\be \label{demando} \left(\int \, \frac{\mathrm{d}x\mathrm{d}
p}{h} \,f \right)^2  \ge \int \, \frac{\mathrm{d}x\mathrm{d}
p}{h} \,f^2. \ee

\subsection{Semiclassical HO-version}
\nd In a semiclassical version, this parameter takes the form \be
\mathcal{D}_{sc}=1-\int\frac{\mathrm{d}^2 z}{\pi}\,\mu(z)^2, \ee
whose analytic expression is \be
\mathcal{D}_{sc}=\frac{1+e^{-\beta\hbar\omega}}{2}. \ee One
ascertains then that for $T \rightarrow \infty$ we have, as
expected, $\mathcal{D}_{sc}=1$. On the other hand, at $T=0$ we get
$\mathcal{D}_{sc}=1/2$.

\subsubsection{\bf{Second echo}}

\nd The above result can be interpreted (\cite{Brazilian} via the
relationship between the decoherence factor and the so-called
participation ratio $\mathcal{R}$, that ``counts" the number of pure states
associated to a density matrix). We find here that just two pure
states
 would ``enter" the semiclassical PDF at $T=0$, if it could be regarded as being of a quantal
 character, since

 \be \label{pratio} \mathcal{D}= 1-\frac{1}{\mathcal{R}}.\ee

\subsection{Escort semiclassical HO-version}
\nd For more interesting results we turn now our attention to
escort distributions in the hope that making $q \ne 1$ may help us
to elucidate more details of our problem. The ensuing
semiclassical version becomes

\be \mathcal{D}_{sc}^{(q)}=1-\int \frac{\mathrm{d}^2 z}{\pi}\,\gamma_q^2,
\ee i.e.,

\be \label{arriba}
\mathcal{D}_{sc}^{(q)}=1-\frac{q}{2}\,(1-e^{-\beta\hbar \omega})=
1-\frac{q}{2}\,\gamma;\,\,\,\,0\le \gamma\le 1. \ee Non-negativity implies
$\frac{q}{2}\,\gamma \le 1$. One can satisfy this relationship and
still retain ample liberty to find acceptable triplets
 of values $\mathcal{D}_{sc}^{(q)}=x,\,q,\,\beta$.
\nd  Additionally, from (\ref{arriba}) we find, calling $x= \mathcal{D}_{sc}^{(q)}$

\be \label{maximo} q= \frac{2(1-x)}{1-e^{-\beta\hbar \omega}}.
\ee

\nd Now, in this case  the Wehrl entropy and Fisher measure turn
out to be, respectively, \cite{Brazilian}

\ben \label{brasil}   &  W_q = 1-
\ln{[q(1-e^{-\beta\hbar\omega})]}\cr & I_q= q (1-e^{-\beta\hbar
\omega}), \een so that the Lieb inequality becomes in this
instance

\ben \label{brasil1} &
 -  \ln{[q(1-e^{-\beta\hbar\omega})]}\ge  0,\,\,i.e., \cr &
  - \ln{q \gamma} \ge 0\,\,\Rightarrow \,\, q \gamma \le 1,\een
which does pose some further constraints on $q$, namely,

\be \label{maximo2} q(1-e^{-\beta\hbar \omega})= 2(1-x) \le 1, \ee
that is

\be \label{maximo1} \mathcal{D}_{sc}^{(q)} \ge 1/2;\,\,\mathcal{R}_{sc}^{(q)}\ \ge 2.\ee

\subsubsection{\bf Third echo}

\nd The meaning of the above result is quite interesting.
Mathematically, $q$ (and thus $I_q$) can be larger than what is
allowed by (\ref{maximo2}), since in such vein one only needs
asking that $W_q\ge 0$, entailing $q \le e/\gamma,$ instead of $q \le
1/\gamma.$ However, for

\be \label{maximo3} 1/\gamma \le q \le e/\gamma, \ee
 Lieb's inequality is violated, which is tantamount to asserting
 that the uncertainty principle is ignored. Thus, we see here that {\it ``too much'' information violates
 Heisenberg's principle} in a semi classical setting.

\subsection{Classical escort version}
\nd The escort classical HO-phase-space probability distribution reads \cite{PPFano}

\be P_q(x,p)=\frac{e^{-q\beta  \hbar\omega |z|^2}}{\int \frac{\mathrm{d} x \mathrm{d} p}{h}\,e^{-q \beta  \hbar\omega |z|^2}},
\ee
so that, after integration one finds
\be
P_q(x,p)=q \beta \hbar\omega \,e^{-q\beta  \hbar\omega |z|^2} \label{clasicalescort}.
\ee

\nd Thus, a simple computation for $\mathcal{D}^{(q)}_{class}=1-\int (\mathrm{d}x\mathrm{d}p/h) P_q(x,p)^2$ yields a result that entails a mere
re-scaling of the inverse-temperature $\beta$ by a factor $q$.

\be \mathcal{D}^{(q)}_{class}=1-\frac{q\beta\hbar\omega}{2}. \ee This
entails a shifting of the minimum allowable energy.

\subsubsection{\bf Fourth echo}

\nd Here  $\mathcal{D}^{(q)}_{class}\geq 0$ entails $q\leq k_B
T/(\hbar\omega/2) $, so that we obtain a physical restriction on
the value of $q$:

\be q\leq \frac{\langle H\rangle_{class}}{E_0}, \ee where
$E_0=\hbar\omega/2$ is the zero-point energy.

\subsection{Quantal escort version}

\nd Interestingly enough, the same $\beta-$rescaling occurs in the
quantum instance. In this version we have $\hat \rho_q=\hat \rho^q/\tr \hat\rho^q=e^{-q\beta \hat H} (1-e^{-q\beta\hbar\omega}) e^{q\beta\hbar\omega/2} $. Thus, the decoherence factor is defined as
$ \mathcal{D}_{quant}^{(q)}=1-\tr \hat \rho_q^2$, and we have the analytic expression

\be \mathcal{D}_{quant}^{(q)}=\frac{2}{1+e^{q \beta\hbar\omega}}. \ee
 We see that $\mathcal{D}_{quant}^{(q)}\geq 0$ implies $q\geq 0$, still
another physical restriction on the $q-$value.

\section{Diverging HO-Fano factors}

\nd It was found in Ref. \cite{PPFMandel} that the semiclassical
q-Husimi-HO treatment reveals the appearance of ``poles", i.e.,
divergences of the Fano factor for specific $q-$values. We delve
further into this issue below.
\subsection{Quantal Fano factor}

\nd If we take  the mean value $\langle \hat H\rangle=\tr (\hat \rho \hat H)$ we have for the quantal Fano factor the expression

\be \mathcal{F}_{quant}=\frac{1}{1-e^{-\beta\hbar\omega}}. \ee For our
present objectives we note that this quantity ``diverges'' {\it only}
for $T=\infty.$

\subsection{Classical Fano factor}

\nd In the classical instance some further consideration become
necessary. The HO's classical partition function was given above
by  $Z_{class}=1/\beta\hbar\omega$
\cite{richard}. Accordingly,

\be\langle \mathcal{ H}\rangle=\frac{\hbar\omega}{Z_{class}}\,\int
(\mathrm{d}x \mathrm{d}p/h)\,|z|^2\,e^{-\beta\hbar\omega
|z|^2}=\frac{1}{\beta}, \ee \be \langle \mathcal{ H}^2\rangle=
\frac{\hbar^2\omega^2}{Z_{class}}\,\int
(\mathrm{d}x \mathrm{d}p/h)\,|z|^4\,e^{-\beta\hbar\omega
|z|^2}=\frac{2}{\beta^2}, \ee which entails $(\Delta
\mathcal{H})^2=1/\beta^2$. As a consequence, we have

\be
\mathcal{F}_{class}=\frac{1}{\hbar\omega\beta-\frac{\hbar^2\omega^2\beta^2}{2}},\label{mandelclasico}
\ee
or

\be
\mathcal{F}_{class}=\frac{1}{\frac{\hbar \omega}{k_B T}-\frac{\hbar^2\omega^2}{2 k_B^2 T^2}}.\label{2mandelclasico}
\ee
At low $T$, $k_B T<<\hbar \omega T$ and $\mathcal{F}_{class}=0.$
 $\mathcal{F}_{class}$ diverges at high temperatures. Indeed, it does so
 at $k_B T=\hbar\omega/2$, when the thermal energy equals the HO-ground state energy.

\subsubsection{\bf Fifth echo}

 \nd This is a quite interesting result.
The classical treatment somehow ``knows'' that this is a strange
energy value, meaningless (but unattainable) in the classical
world, and reacts with a ``pole''. In any case, classical
considerations do lead to the vacuum HO-energy (again!).

\subsection{Semiclassical Fano factor}
%%%%%%%%%%%%%%%%\subsection{Presentation}
%%%%%%%%%%%%%%%%%%%%%%%%%%%%%%%%%%%%%%%%%%%%

 \nd The semiclassical version $\mathcal{F}_{sc}$ of Fano factor evaluated with  Husimi's distribution was found in Refs.~\cite{PPFano,PPFMandel}

\be \mathcal{F}_{sc}=\frac{(\Delta_{\mu} N)^2}{\langle \hat
N\rangle_{\mu}}, \ee where  $\langle \,...\, \rangle_\mu$
denotes the semiclassical mean value of  any general observable
and the subindex $\mu$ indicates that we have taken the
 Husimi distribution (\ref{mu1}) as the weight
function. It is then easy to see that $\mathcal{F}_{sc}$ reads  \be \mathcal{F}_{sc}=
\frac{2}{
(1-e^{-\beta\hbar\omega})(2-(1-e^{-\beta\hbar\omega}))}. \label{why}
 \ee
No divergences ensue in this instance. However, they will appear
if we appeal to escort distributions.

\section{Escort Fano factors}
\subsection{Semiclassical escort Fano factor for the HO}

\nd  The ``escort"-expression for the  Fano factor is~\cite{PPFMandel}

\be \mathcal{F}_{sc}^{(q)}=\frac{2}{q
(1-e^{-\beta\hbar\omega})(2-q(1-e^{-\beta\hbar\omega}))}.  \ee We
note that when $q$ tends to unity  we have $\mathcal{F}_{sc}^{(1)}\equiv
\mathcal{F}_{sc}$. We see now the Fano-divergences may occur whenever

\be \label{alfa} \frac{2}{q}= G(\beta) = 1-e^{-\beta\hbar\omega}.    \ee
Since $0 \le \exp{(-\beta\hbar\omega)} \le 1$

\be \label{alfa1} 0 \le G(\beta) \le 1,   \ee and \be
\label{0alfa2} 2 \le q \le \infty.   \ee Additionally, the inverse
temperature at which the divergence of the Fano factor takes place
is given by

\be \label{alfa2}  \beta_{Fdiverg}(q)=
\frac{-\ln{(1-2/q)}}{\hbar\omega}, \ee a value that obviously
ranges in $[0,\infty]$. We conclude that the ``classical pole'' can
be ``moved'' to any temperature whatsoever by a judicious choice of $q$, which allows one then to
mimic at will the ``pole''-behavior in either the classical or
the quantum (at $T= \infty$) instances.

\subsubsection{\bf Second observation}

\nd The escort distribution can mimic, after judicious
$q-$selection and for specific physical facets,  either quantum or
classic behavior.

\subsection{Escort-classical Fano factor}

\nd The escort-classical HO-phase-space probability distribution found in (\ref{clasicalescort}) that reads $P_q(x,p)=q \beta \hbar\omega \,e^{-q\beta  \hbar\omega |z|^2}$, and using  $\langle f \rangle=\int (\mathrm{d} x \mathrm{d} p/h)\,f(x,p) P_q(x,p)$ one obtains
 $\langle H\rangle=1/(q\beta )$, $\langle H^2\rangle=2/(q^2 \beta^2)$, and $(\Delta H)^2=1/(q^2\beta^2)$.
Consequently, the $q$-escort classical  Fano factor is

\be
\mathcal{F}_{class}^{(q)}=\frac{1}{q \beta \hbar \omega-q^2 \beta^2 \frac{\hbar^2\omega^2}{2}} \label{qmandelclasico},
\ee
or

\be
\mathcal{F}_{class}^{(q)}=\frac{1}{q \frac{\hbar \omega}{k_B T}-q^2 \frac{\hbar^2\omega^2}{2 k_B^2 T^2}} \label{2qmandelclasico}.
\ee The  limit $q\rightarrow 1$ leads to $\mathcal{F}_{class}^{(1)}\equiv \mathcal{F}_{class}$.
The Fano ``pole'' becomes located at $q=2 k_B T/(\hbar\omega)$. Also, here we have

\be \label{newalfa}   \beta_{Fdiverg}(q)= \frac{2}{q\hbar\omega}, \ee and can be chosen at will.
%%In Fig. \ref{mandelclasicoT} we observe several curves. Small-dashed for $q=1$, black for $q=0.5$ and, large-dashed for  %%$q=1.5$.
%%The net escort-effect is a mere change of the temperature-scale, from $T$ to  $q\,T$.
%%% However, this suffices to locate the pole anywhere.

%\begin{figure}[ht]
%     \begin{center}
%       \includegraphics[scale=0.38,angle=0]{1ClasicoFanoT.eps}
%       \caption{$\mathcal{F}_{class}^{(q)}$  as a function of temperature $T$ in $\hbar\omega/k_B-$units for several values of %%%$q$.}\label{mandelclasicoT}
   %      \end{center}
 %%%%%%%%%%%%%%%\end{figure}

\subsection{Quantal escort-Fano factor} \nd   Here we have

\be
\mathcal{F}_{quant}^{(q)}=\frac{1}{1-e^{-q \beta \hbar \omega}},
\ee
i.e., we find again a $q\beta-$scaling and nothing interesting happens. $e^{-q \beta \hbar \omega}=1$
when either $q=0$ or  $T\rightarrow \infty$.

%%%%%%%%%%%%%%%%%%%%%%%%%%%%%%%%%%%%%%%%%
\section{Conclusions}\label{remarks}
%%%%%%%%%%%%%%%%%%%%%%%%%%%%%%%%%%%%%%%%%

\nd We have focused attention here on two concepts: the
decoherence parameter $\mathcal{D}$ and the divergence of the Fano
factor for specific $q$ or $\beta$ values. These two notions have
been treated at three levels: 1) quantum, 2) classical, and 3)
semiclassical. In all instances this was done both for  $q=1$ and
$q \neq 1$.

\nd We have heard quantum echoes at the classical level and
discovered that by changing $q$ we can force the semiclassical
results to accommodate either quantum or classical properties.

\nd  In related matters concerning stochastic electrodynamics, the
illuminating work of T.H. Boyer and L de la Peña et al. (among
others) has to be mentioned \cite{boyer,penha,cetto}, what we here call  echoes emerge
there as well. It is safe then to assert then that the
classical-quantum links deserve further scrutiny.

\section*{Acknowledgements}
 F. Pennini would like to thank partial financial support by FONDECYT, grant 1110827.

%%%%%%%%%%%%%%%%%%%%%%%%%%%%%%%%%%%%%%%%%%%%%%%%%%%%%%%%%%%%%%%%%%%%%%%%%%%

\end{document}